\documentclass{elsart}
\usepackage{graphics}
\usepackage{graphicx}
\usepackage{natbib}
\usepackage{amssymb}
\def\farcs{\hbox{$.\!\!^{\prime\prime}$}}

\def\facs{\hbox{$.\!\!^{s}$}}

\def\h2o{H$_2$O}
\def\kms{km~s$^{-1}$}
\def\hi {H\kern0.1em{\sc i}}
 \def\hii {H\kern0.1em{\sc ii}}

\def\kms{km s$^{-1}$}

\def\solum {\hbox{L$_{\odot}$}}

\begin{document}
\begin{frontmatter}


\title{The water megamaser in the merger system Arp~299}
\author[oac,ira]{A.\ Tarchi\corauthref{cor}},
\corauth[cor]{Corresponding author.}
\ead{atarchi@ca.astro.it} 
\author[oac]{P.\ Castangia},
\author[mpi]{C.\ Henkel}, and 
\author[mpi]{K.\ M.\ Menten}

\address[oac]{INAF - Osservatorio Astronomico di Cagliari, Poggio dei Pini, Strada 54, 09012 Capoterra (CA), Italy}
\address[ira]{INAF -- Istituto di Radioastronomia, via Gobetti 101, 40129 Bologna, Italy}
\address[mpi]{Max-Planck-Institut f\"{u}r Radioastronomie, Auf dem H\"{u}gel 69, 53121 Bonn, Germany}

\begin{abstract}
We present preliminary results of an interferometric study of the water megamaser in the merger system Arp~299. This system is composed of two main sources: IC~694 and NGC~3690. There is clear evidence that most of the water maser emission is associated with the nucleus of the latter, confirming the presence of an optically obscured AGN as previously suggested by X-ray observations.
Furthermore, emission arises from the inner regions of IC~694, where an OH megamaser is also present. The velocity of the water maser line is blueshifted w.r.t. the optically determined systemic velocity and is consistent with that of the OH megamaser line. This finding might then indicate that both masers are associated with the same (expanding) structure and that, for the first time, strong 22\,GHz \h2o\ and 1.67\,GHz OH maser emission has been found to coexist.
\end{abstract}
\end{frontmatter}

\section{Introduction}
\label{intro}
Luminous sources of extragalactic \h2o\footnote{In the following, when talking about \h2o\ masers, we will only refer to the $6_{16} - 5_{23}$ transition of \h2o (rest frequency: 22.23508\,GHz). Recently, however, luminous extragalactic maser emission has been detected also from the $3_{13} - 2_{20}$ transition of \h2o (rest frequency: 183.310\,GHz) in the starburst galaxy NGC~3079 (Humphreys et al.\ 2005) and in the ultra-luminous infrared OH megamaser galaxy Arp~220 (Cernicharo et al.\ 2006).} and OH emission, the `megamasers' (with isotropic luminosities $L_{\rm iso}$$>$10\,L$_{\odot}$), are known to arise from the highly obscured inner parsecs of galaxies which host an active nucleus (AGN) or from the dense interstellar medium of UltraLuminous InfraRed Galaxies (ULIRGs). Interferometric studies of such \h2o\ masers provide the only known probe to map molecular tori and accretion disks and to study their geometry and thickness (e.g. Miyoshi et al.\ 1995), while OH can trace more extended structures in merging spirals (e.g. Kl\"ockner et al.\ 2003). 

\h2o\ and OH masers can also be associated with star forming regions and oxygen-rich late-type stars, but in these cases their luminosities are much weaker than those of the megamasers. The most extreme case with `weaker' \h2o\ emission known to date, the starburst spiral galaxy NGC~2146, shows two clusters of \h2o\ masers, each with a luminosity of $\sim$1\,L$_\odot$ and a total single-dish luminosity of $\sim$8\,L$_\odot$, presumably all related to massive star formation (Tarchi et al.\ 2002).

Although the two most prominent extragalactic maser species, \h2o\ and OH, are associated with similar objects, a coincidence in the same object has only been found in M~82, NGC~3079, and NGC~4945, and has never involved particularly luminous maser sources from both molecular species. Hence, except for these galaxies, \h2o\ and OH emission seem to be mutually exclusive. This is likely a consequence of the diverse excitation and pumping conditions of the respective lines. OH maser emission, excited mostly radiatively and requiring comparatively low density gas, is more extended and farther off the excitation source than the water masers that are likely collisionally excited and require higher temperature and density. As described in the following, we may have found a first target where megamaser from both species are detected. 

\section{The Arp~299 system}
The extremely luminous infrared galaxy Arp~299 (Mkn~171)\footnote{According to the NASA/IPAC Extragalactic Database (NED), the entire merger forms NGC 3690, while IC 694 is a less prominent galaxy 1$^{\prime}$ to the northwest. Here we follow the more traditional nomenclature that is commonly used in the literature.} is a merging system located at a distance of $\sim$42\,Mpc. It is composed of four main regions of activity: the two galaxies IC~694 ({\bf A}) and NGC~3690 ({\bf B}; resolved at infrared and radio wavelengths into two components, {\bf B1} and {\bf B2}), and two individual concentrations ({\bf C} and {\bf C$^{\prime}$}) at the interface where IC~694 and NGC~3690 overlap (e.g. Neff et al.\ 2004; Fig.~1; large panel). The system is rich in molecular gas (Casoli et al.\ 1999) and displays OH megamaser activity with an isotropic luminosity of $\sim$ 240 \solum (Baan 1985) apparently tracing a rotating disk in IC~694 (Baan \& Haschick 1990; hereafter BH90). The presence of such a flattened rotating structure in IC~694 was subsequently also invoked by Polatidis \& Aalto (2001) to explain its atomic (\hi) and molecular (CO) gas velocity distribution. In March 2002, strong water megamaser emission was detected in the merging system Arp 299, with a total isotropic luminosity of $\sim$ 200 L$_\odot$ (Henkel et al.\ 2005).

\section{Observations and data reduction}
Water vapor in Arp~299 was observed on September 19, 2004, with the Very Large Array (VLA\footnote{The National Radio Astronomy Observatory (NRAO) is operated by Associated Universities, Inc., under a cooperative agreement with the National Science Foundation.}) in its A configuration and a frequency setup with two 25 MHz intermediate frequency bands (``IFs'') centered at Local Standard of Rest velocities of $V_{\rm LSR}$ = 2900 and 3200 \kms. The IFs were overlapped in frequency in order to minimize the effect of band-edge roll-off.
Each IF covered $\sim$ 340 $\rm km\:s^{-1}$ with a resolution of $\sim$ 20 $\rm km\:s^{-1}$. The data were Fourier-transformed using natural weighting to create a data cube. A radio continuum map was produced using the line-free channels. The restoring beam was 0\farcs1 $\times$ 0\farcs1 and the rms noise was 0.5 mJy/beam/chan and 0.1 mJy/beam for the cube and the continuum maps, respectively. The noise was higher than the expected thermal noise, likely because of poor weather during the observations. 

\section{Preliminary results and discussion}
\begin{figure}
\begin{center}
\includegraphics[width=13cm]{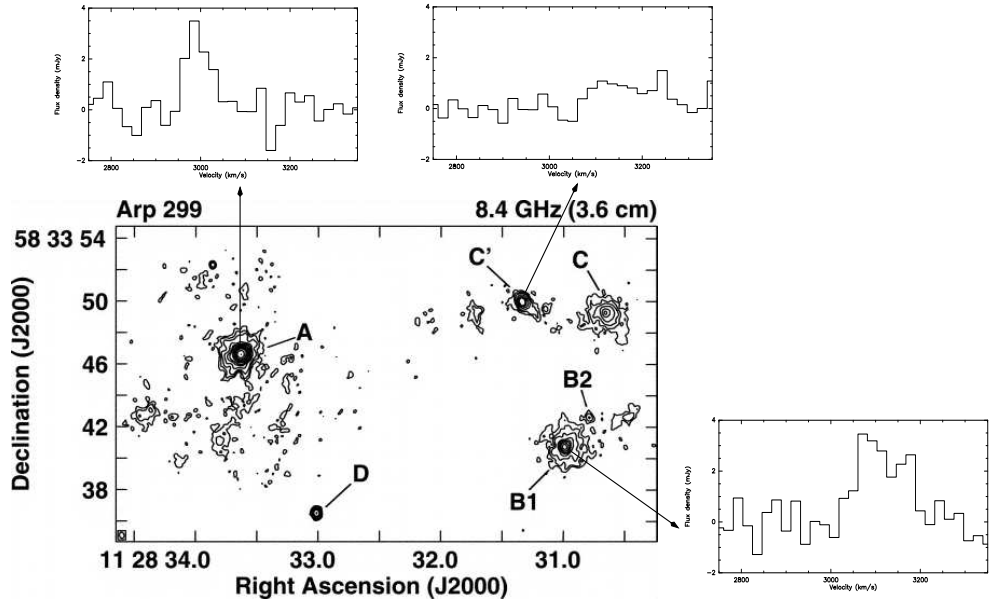}
\end{center}
\caption{A VLA 8.3 GHz continuum map of Arp~299 at a resolution of $\sim$ 0\farcs4 (FWHM). For details, see Neff et al.\ 2004. In the small panels, water maser spectra at three locations (corresponding to the nuclei of IC~694 and NGC~3690 and the overlapping region C$^{\prime}$) taken with the VLA A-array at a resolution of 0\farcs1.}
\label{bigone}
\end{figure}

 Fig.~\ref{bigone} shows a composition of a VLA 8.4 GHz continuum map of Arp~299 at a resolution of $\sim$ 0\farcs4 (FWHM) (big panel; for details, see Neff et al.\ 2004) and the water maser spectra (small panels) at three locations (corresponding to the nuclei of IC~694 and NGC~3690 and the overlapping region C$^{\prime}$) observed by us with the VLA A-array at a resolution of 0\farcs1. The \h2o\ emission originates from more than one hotspot. Most of the emission is seen associated with the radio continuum nucleus of NGC~3690 at position RA${\rm _{J2000}}$ = 11$^{\rm h}$ 28$^{\rm m}$ 30\facs99; Dec${\rm _{B2000}}$ = 58$^{\rm \circ}$ 33$^{\rm \prime}$ 40\farcs7. This maser feature seems to be composed of two subcomponents with V$_{\rm LSR}$ = 3070 and 3170 \kms, respectively. With a peak flux density of $\sim$ 4 mJy/beam and a global width of $\sim$ 150 \kms, the feature has an isotropic luminosity of 25 \solum. Furthermore, emission arises also from the inner regions of the galaxy IC~694 at position RA${\rm _{J2000}}$ = 11$^{\rm h}$ 28$^{\rm m}$ 33\facs65; Dec${\rm _{B2000}}$ = 58$^{\rm \circ}$ 33$^{\rm \prime}$ 46\farcs8. This feature, at V$_{\rm LSR}$ = 2985 \kms, has a peak flux density of $\sim$ 4 mJy/beam and a width of $\sim$ 80 \kms, and hence, an isotropic luminosity of 12 \solum. The error in the absolute positions with detected emission is conservatively assumed to be the VLA nominal one, i.\ e.\ 0\farcs1 ($\sim$ 20 pc). Emission is tentatively detected as well in the overlapping region at the position RA${\rm _{J2000}}$ = 11$^{\rm h}$ 28$^{\rm m}$ 31\facs24; Dec${\rm _{B2000}}$ = 58$^{\rm \circ}$ 33$^{\rm \prime}$ 52\farcs8. Being observed at a 2$\sigma$ level, this result has to be confirmed by new observations and is therefore not discussed further.

\subsection{The origin of the maser emission.}

For many years, the Arp~299 complex was investigated in a search for a signature of AGN activity in the central regions of NGC~3690 and IC~694. While the presence of highly-absorbed but intrinsically strong X--ray emission was detected in Arp~299 at E $>$ 10 keV (Della Ceca et al.\ 2002) suggesting the presence of a luminous buried AGN, the spatial resolution was not good enough to determine the location of the putative AGN within Arp~299. More recent higher spatial resolution X--ray data strongly support the presence of a buried AGN in the mid-infrared peak B1 of NGC~3690 (Zezas et al.\ 2003; Ballo et al.\ 2004). The association of the component B1 with the true (active) nucleus of NGC~3690, has been supported further using mid-infrared (Gallais et al.\ 2004) and optical (Garcia-Marin et al.\ 2006) spectroscopy. Within this framework, our detection of a megamaser (although weaker than the single-dish detection, see next section) coincident in position with the spot B1 constitutes an additional proof of the presence of an AGN in NGC~3690. The velocity of the maser line is also consistent with that of the molecular gas (CO) in NGC~3690, when taking into account the complexity of the velocity field of the system (Casoli et al.\ 1999).

A less straightforward scenario is instead offered by the detection of maser emission within 0\farcs1 ($\sim$ 20 pc) of the nucleus of IC~694. The presence of a second active nucleus in the interacting system Arp~299 is still under discussion. All the observations (radio, infrared, and X-rays) offer an ambiguous interpretation that allows to support either the AGN nature of the nucleus of IC~694 (e.g. Ballo et al.\ 2004) or the presence of a deeply embedded nuclear starburst (e.g. Gallais et al.\ 2004). The presence of an OH megamaser has been interpreted as due to low-gain amplification of a diffuse underlying background continuum structure possibly constituted by a nuclear starburst (BH90). The isotropic luminosity of the water maser feature detected by us is just above the 10 \solum\ threshold conventionally used to discriminate between the more luminous megamasers (typically indicative of AGN) and the weaker kilomasers (mostly associated with enhanced star formation activity), and hence, it cannot be confidently used to discriminate between the two scenarios.

The velocity of the \h2o\ maser line can, however, be used to obtain a clearer picture. The maser velocity is blueshifted by $\sim$ 100 \kms\ w.r.t. the optical estimate of the systemic velocity of IC~694 (V$_{\rm sys}$ = 3100 -- 3110 \kms; BH90 and references therein). CO, \hi\, and OH emission detected in IC~694 show also components with velocities smaller than the systemic one and are consistent with that of the maser line (Casoli et al.\ 1999; BH90). BH90 suggested that the blueshifted \hi\ and OH gas is part of an expanding structure in the foreground of the nucleus of IC~694. The expanding structure could be made up from material swept up by an episode of nuclear activity, possibly representing a nuclear (AGN or starburst driven) outflow. The association of luminous water maser emission with an (AGN) outflow has been so far reported only for the Circinus galaxy (Greenhill et al.\ 2003), and hence, if confirmed, our finding might represent a second important case that needs, in our opinion, to be investigated in more detail. It is remarkable that IC~694 represents the first case where strong \h2o\ and OH maser emission have been detected in the same galaxy. 

\subsection{The missing flux problem}
\begin{figure}
\begin{center}
\includegraphics[width=8cm]{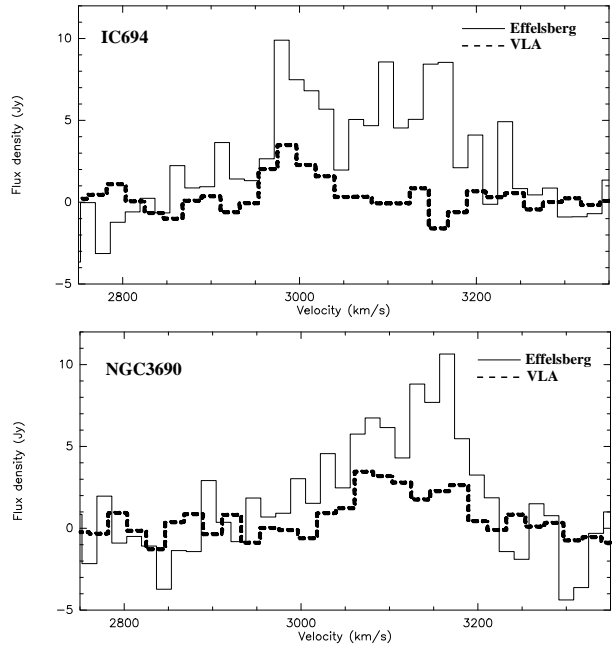}
\end{center}
\caption{\h2o\ maser profiles toward the nuclei of IC~694 ({\it top panel}) and NGC~3690 ({\it bottom panel}), taken with the 100-m telescope at Effelsberg (November 2005; solid lines) and the VLA A-array (September 2004; dashed lines). The channel spacing is  17 and 20 km\,s$^{-1}$, respectively.}
\label{comp}
\end{figure}

The maser features detected with the VLA in NGC~3690 and IC~694 have velocities consistent with that of the (broader) feature observed with Effelsberg\footnote {The 100-m telescope at Effelsberg is operated by the Max-Planck-Institut für Radioastronomie (MPIfR) on behalf of the Max-Planck-Gesellschaft (MPG).} in 2002. The flux density peak is, however, significantly degraded (from 25 to 4 mJy/beam) w.r.t. the Effelsberg spectra (see Henkel et al.\ 2005; their Fig.~6). This fact is readily explained in two ways:
\begin{itemize}

\item since variability is a common phenomenon among maser lines, the flux loss could be just an intrinsic dimming. To test this option, we have compared the VLA spectra with two more recent (November 2005) Effelsberg observations pointed toward the two nuclei of the galaxies in Arp~299 (Fig.~\ref{comp}). As before, the single-dish lines are much stronger than those measured by the VLA and do not  support strong variability.

\item the maser emission may consist of compact hotspots spread over a wide area and the two features detected may represent only the strongest centers of emission (see Sect.~1 for a similar situation in NGC~2146). VLA 22~GHz high-sensitivity observations with angular resolutions of 1--3 arcseconds have been requested to study this possibility and to locate a larger fraction of the total H$_2$O flux.
\end{itemize}


\section{Concluding remarks} 

We have mapped the 22~GHz water vapor maser emission in Arp~299 with the VLA in its most extended configuration. Our preliminary analysis indicates that two maser spots are associated with the nuclear regions of NGC~3690 and IC~694. Another tentatively detected weaker spot is possibly associated with the overlapping region. Given these results, we conclude that: 1) The presence of an AGN in NGC~3690 is confirmed; 2) IC~694 represents the first case where luminous \h2o\ and OH masers are coexisting; 3) the velocity of the maser line in IC~694 is blueshifted w.r.t. the systemic velocity of the galaxy indicating an association with an expanding structure previously seen in \hi\ and OH; 4) our results on IC~694 are not incompatible with the presence of a second AGN in Arp~299  

\noindent{\it {Acknowledgements}}

AT wishes to thank Willem Baan and Susanne Aalto for useful discussions. This research has made use of the NASA/IPAC Extragalactic Database (NED) which is operated by the Jet Propulsion Laboratory, California Institute of Technology, under contract with the National Aeronautics and Space Administration.

\end{document}